# Integrated Approach of Gearbox Fault Diagnosis

Vikash Kumar[1], Subrata Mukherjee[2][0000-1111-2222-3333] and Somnath Sarangi[1]

[1]Indian Institute of Technology Patna, Patna, Bihar 801106, India
[2]Institute of Engineering & Technology, Patiala, Punjab 147004, India
{ vikashkumar7r,subrata.re.me}@gmail.com
somsara@iitp.ac.in

**Abstract.** Gearbox fault diagnosis is one of the most important parts in any industrial systems. Failure of components inside gearbox can lead to a catastrophic failure, uneven breakdown, and financial losses in industrial organization. In that case intelligent maintenance of the gearbox comes into context. This paper presents an integrated gearbox fault diagnosis approach which can easily deploy in online condition monitoring. This work introduces a nonparametric data preprocessing technique i.e., calculus enhanced energy operator (CEEO) to preserve the characteristics frequencies in the noisy and inferred vibrational signal. A set of time domain and spectral domain features are calculated from the raw and CEEO vibration signal and inputted to the multiclass support vector machine (MCSVM) to diagnose the faults on the system. An effective comparison between raw signal and CEEO signal are presented to show the impact of CEEO in gearbox fault diagnosis. The obtained results of this work look very promising and can be implemented in any type of industrial system due to its nonparametric nature.

## 1  Introduction

Gearbox system is the one of the most important key elements in any mechanical system because of its compact mechanical structure with huge power transmission capacity. However, gearboxes are used in extremely complex industrial environments with high impulse load, cyclic loading and stress conditions which further lead to the failure of component inside the gearbox system [1-3]. Therefore, a safe fault diagnosis of the gearbox system is a prime necessity for intelligent maintenance in industrial systems.

Fault diagnosis of the gearbox system can be done based on the vibration [1-4], current [5-6], acoustic or oil-based signal analysis etc. [2,7]. Among them vibration signal-based analysis is extremely sensitive to any type of gearbox faults and also, it is widely used techniques for gearbox fault diagnosis. Gearboxes are operated under harsh environmental condition such as sudden shock load, high compressive contact stress, improper lubrication etc. due to which the components inside gearbox start fail over a period of time. This increases the vibration level of the system and at the time of data acquisition the other non-stationary properties like noise and interferences indulge with the acquired signal and affect its quality [3,7,8]. Many research works have already been published regarding the separation of these non-stationary properties from the vibration signal but it still become the area of research in industry sector.

Numerous authors presented the empirical study on vibration-based signal processing methods for mechanical system failure such as time domain analysis, frequency domain analysis, and time-frequency analysis [1,3,8,9]. But all these methods have some limitations such as requirement of prior knowledge of system in TSA [10,11], appropriate bandpass filter design [12], window resolution problem in STFT [12], cross-term problem in WVD [12], high time requirement for analysis in CWD [12], proper selection of wavelet parameters in wavelet transform [12] etc. that may become the obstacles in online fault diagnosis process.

Therefore, for real time fault diagnosis, there is a need to develop a technique that is independent of prior knowledge of the system, nonparametric, non-filtering, easy to implement, and computationally efficient. In this paper, a nonparametric data pre-processing (CEEO) is integrated with the classifier MCSVM for fault diagnosis of gearbox. The proposed technique has capability to work well under non-stationary elements.
CEEO helps to preserve the important frequencies such as pinion frequency, gear frequency, gear mesh frequency and its harmonics, sidebands of modulating frequencies etc. inside the vibration signal even in the presence of strong noise and multiple vibration interferences. It not only helps to build better quality of feature generation in different domain such as time domain, frequency domain, time-frequency domain etc. but also plays a major role in improving the classification accuracy of the classifier. To show the effectiveness of the proposed technique the results are compared with and without CEEO data pre-processing. Also, importance of CEEO is shown by comparing the results by adding spectral features to the time domain features.

Rest of the paper is summarized in the following way: Section 2 describes the brief theoretical study of the CEEO and MCSVM with proposed methodology, Section 3 presents the full details of the experiential framework with vibration signal processing, Section 4 analyses the preliminary outcomes of the proposed method, and Section 5 demonstrates the conclusion and future work.

## 2  Theoretical Backgrounds and Methodology

### 2.1  Calculus Enhanced Energy Operator (CEEO)

CEEO is the modified version of Energy Operator (EO) [12] and based on layer operator (LO). The LO incorporates the sequential operation of both differentiation and integration elements either in forward or backward direction to cope the limitation of EO [12]. The EO is sensitive to noise but at the same time it is very effective in preserving the impulse component in the signal by improving signal to interference ratio (SIR) [13,14]. In CEEO, differentiation operator helps to improve the SIR and integration operator helps to improve the signal to noise ratio (SNR) in the signal [12]. These properties of CEEO help to generate good quality of signal that further enhance the quality of extracted features. The discrete form of CEEO is calculated in the same way as the EO by replacing the 1$^{st}$ and 2$^{nd}$ order derivatives of signal with 1$^{st}$ and 2$^{nd}$ order LO and is defined as [12]:

$$CEEO\left(x(n)\right) = LO_1^2\left(x(n)\right) - x(n)LO_2\left(x(n)\right) = x^2(n) - x(n-2)x(n+2) \quad (2)$$

### 2.2 Multi Class Support Vector Machine (MCSVM)

SVM was developed by Vapnik [15] to solve the regression and classification problem. Initially, it was designed to solve two class classification problem. But, due to better generalization and ability to produce high classification accuracy with limited datasets, it is extended to solve multi-class classification problem. In general, two approaches to multi-class SVM are commonly used: "one-versus-all" and "one-versus-one". For N class classification, one-vs-all build N binary model in such a way that the sample data from N$^{th}$ model are assigned as *+ve* class and rest sample data are assigned as -ve class while one-vs-one build N(N-1)/2 binary model and choose only two class at a time, it takes sample data from these classes to train the binary model by ignoring the sample data from rest classes. A voting strategy is applied among the models in both cases to predict the class of test sample data [15-17]. Hsu et. al [16] have studied different types of SVM classifier with different multi class problem datasets and found that one-vs-one SVM classifier has more accuracy to classify multi class problem than other types of SVM classifier.

### 2.3 Proposed Methodology

Figure 1 shows the proposed technique for the fault diagnosis of gearbox. Firstly, Experimental vibration data are acquired from the test rig shown in Figure 2. The acquired data are pre-processed by CEEO to build good quality of signal. Various time domain and frequency domain features are extracted from the CEEO processed data. After that, both features are collectively input to the one-vs-one MCSVM classifier for fault diagnosis of gearbox.

## 3 Experimental Setup and Data Collection

All the experimental work were conducted on the machine fault simulator (MFS) as shown in Figure 2. Overall experimental setup comprises four major subsystems: MFS, Oros NV gate data acquisition system, various sensors, and a computer, as shown in Figure 2. In this work, the main experimental investigations were performed on a single-stage bevel gearbox which has 18 teeth in driver gear and 27 teeth in the driven gear. Figure 3 shows the faulty condition of gear tooth which were used in the experimental analysis. Samples were recorded at different operating condition such as by varying the frequency of motor (15 Hz, 25 Hz, 35 Hz), load on gearbox (0 lb., 2 lb., 4 lb.)

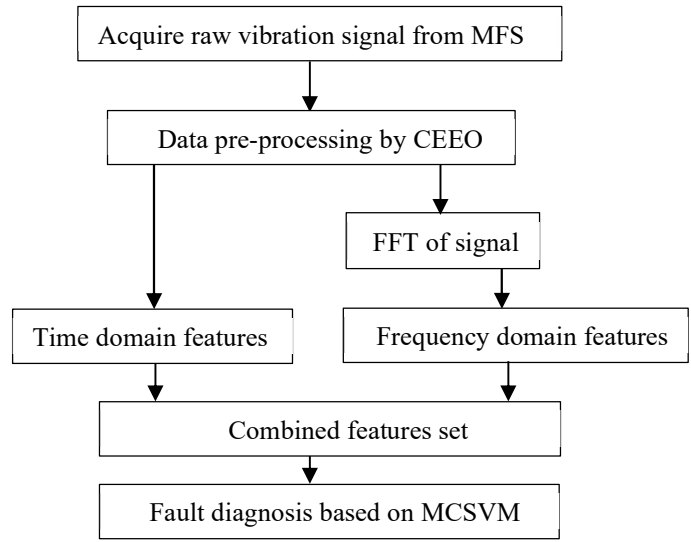

**Fig. 1.** Proposed fault diagnosis technique for gearbox

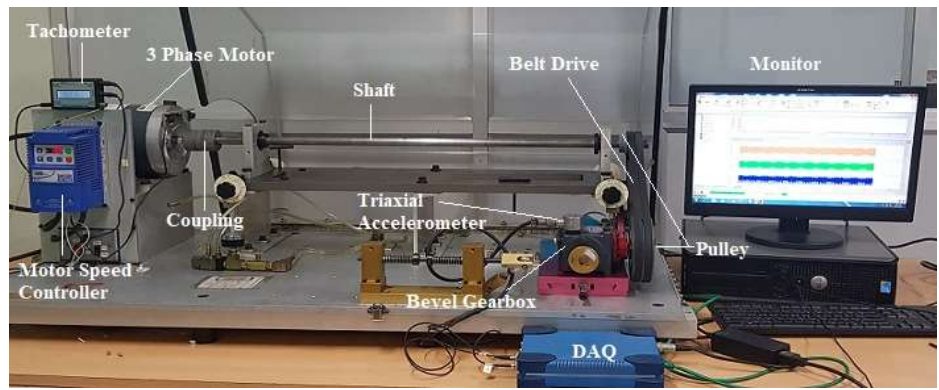

**Fig. 2.** Experimental setup for data acquisition.

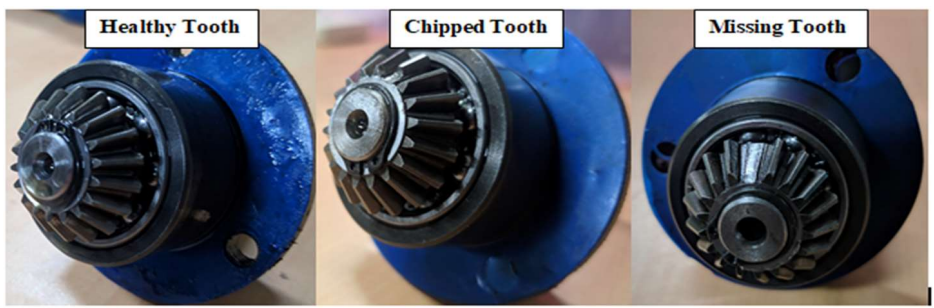

**Fig. 3.** Types of gear faults on gearbox.

for three health condition of gear tooth (Healthy, Chipped, Missing). Total of 27 samples were collected in which each 9 samples were from each fault condition cases. The sampling frequency of the acquired vibration signal was set at 2048 samples/sec and each sample was record for 30 sec.

## 4    Results and Discussion

### 4.1    Data Pre-Processing

All collected samples are pre-processed by CEEO. Figure 4 shows the time and frequency plot of raw signal at 25 Hz motor frequency and 0 lb. load. Figure 5 shows the same for processed (CEEO) signal. Figure 4(a) and 5(a) clearly show linear trend between the amplitude of vibration response and severity of fault on the system. The missing tooth has higher response amplitude than the chipped and healthy condition. Figure 4(b) and 5(b) show the FFT plot of vibration signal shown in Figure 4(a) and 5(a), respectively. From the FFT plots, it can be seen that the suppress frequencies such as pinion, gear and its harmonics are more accurate and visible in the FFT plot of CEEO i.e., Figure 5(b). Also, the sidebands of faulty frequencies (modulating frequency of pinion) are clearly visible around the gear mesh frequency (GMF) and its harmonics. Figure 5(b) clearly shows the growing nature of these frequencies as severity of fault increases in the system while Figure 4(b) shows the mixed behaviors due to the presence of non-stationary elements in the signal. All these properties show the effectiveness of CEEO in handling the non-stationary elements in the signal. These properties help in improving the quality of signal and produce good quality of features that further improve the classification accuracy of classifier.

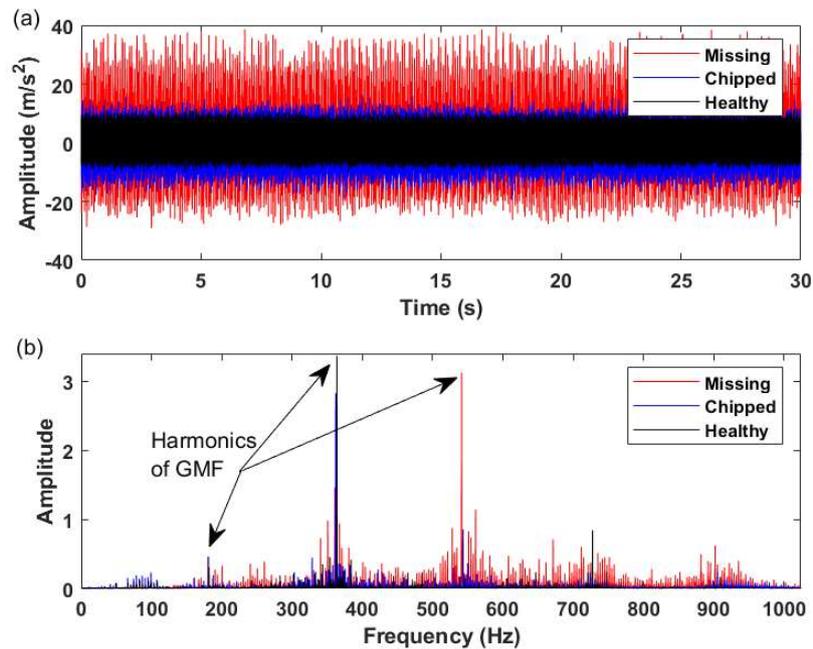

**Fig. 4.** Raw signal at 25 Hz motor frequency and 0 lb. load (a) Vibration response (b) FFT

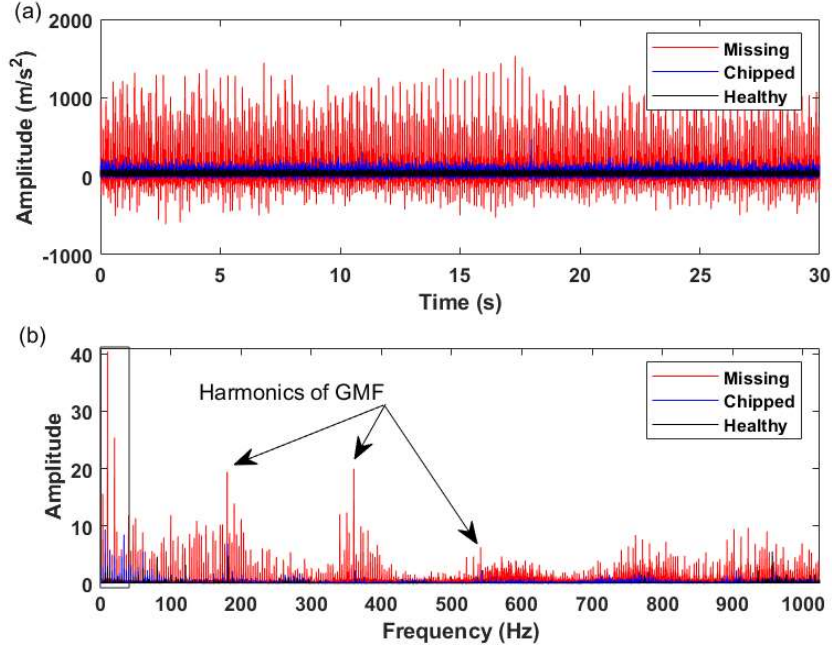

**Fig. 5.** CEEO signal at 25 Hz motor frequency and 0 lb. load (a) Vibration response (b) FFT.

### 4.2 Feature Extraction

In this work, total 19 features (12-time domain and 7 frequency domain) are extracted from the raw signal and CEEO signal, presented in Table 1 and Table 2 respectively. Formulation of these features are given in ref. [1,18]. These features have their own importance to predict the different types of behaviors from the signal, which play an important role in fault classification. For example, time domain features like peak value detects the severity of faults, RMS detects the system's unbalanced parts, standard deviation detects the randomness of non-stationary system, crest factor measures the spikiness or existence of sharp peak in the signal, skewness i.e., third order moment of the signal indicates the position or orientation of defects, kurtosis i.e., fourth order moment of the signal identifies the worn and damage tooth of gears. Also, frequency domain features have some special ability to detect the gearbox faults, same as the time domain features [18].

**Table 1.** Time domain features formulation

| Feature | Expression |
|---|---|
| Peak | $\max\lvert Y(n)\rvert$ |
| Mean ($\mu$) | $\dfrac{\sum_{n=1}^{N} Y(n)}{N}$ |
| RMS | $\sqrt{\dfrac{\sum_{n=1}^{N} Y(n)^2}{N}}$ |
| Crest factor | $\dfrac{Peak}{RMS}$ |
| Impulse factor | $\dfrac{Peak}{1/N \sum_{n=1}^{N} \lvert Y(n)\rvert}$ |
| Clearance factor | $\dfrac{Peak}{\left(1/N \sum_{n=1}^{N} \sqrt{\lvert Y(n)\rvert}\right)^2}$ |
| Varience | $\dfrac{\sum_{n=1}^{N} (Y(n) - \mu)^2}{N - 1}$ |
| Std. deviation ($\sigma$) | $\sqrt{Varience}$ |
| Skewness | $\dfrac{\sum_{n=1}^{N} (Y(n) - \mu)^3}{(N - 1)\sigma^3}$ |
| Kurtosis | $\dfrac{\sum_{n=1}^{N} (Y(n) - \mu)^4}{(N - 1)\sigma^4}$ |
| SNR | $10 \log_{10}\left(\dfrac{P_{signal}}{P_{noise}}\right)$ |
| SINAD | $10 \log_{10}\left(\dfrac{P_{signal}}{P_{noise} + P_{distortion}}\right)$ |

Where $Y$ denotes the vibration signal and N is the total samples in the signal

**Table 2.** Spectral features formulation

| Feature | Expression |
|---|---|
| Spectral Centroid ($\mu_s$) | $\dfrac{\sum_{n=1}^{N} (f(n) \times F(n))}{\sum_{n=1}^{N} F(n)}$ |
| Spectral Spread ($\sigma_s$) | $\sqrt{\dfrac{\sum_{n=1}^{N} ((f(n) - \mu_s)^2 \times F(n))}{\sum_{n=1}^{N} F(n)}}$ |
| Spectral Skewness | $\dfrac{\sum_{n=1}^{N} (f(n) - \mu_s)^3 \times F(n)}{\sigma_s^3 \times \sum_{n=1}^{N} F(n)}$ |
| Spectral Kurtosis | $\dfrac{\sum_{n=1}^{N} (f(n) - \mu_s)^4 \times F(n)}{\sigma_s^4 \times \sum_{n=1}^{N} F(n)}$ |
| Spectral Entropy | $\dfrac{-\sum_{n=1}^{N} F(n) \times \log F(n)}{\log(length(F(n)) - 1)}$ |
| Spectral Crest | $\dfrac{\max F(n)}{1/(length(F(n)) - 1) \times \sum_{n=1}^{N} F(n)}$ |
| Spectral Slope | $\dfrac{\sum_{n=1}^{N} [(f(n) - mean(f(n))) \times (F(n) - \mu_s)]}{\sum_{n=1}^{N} [f(n) - mean(f(n))]^2}$ |

Where $F$ is the absolute value of spectral signal of the corresponding time-domain signal *(Y)* in Table 1, $f = fs\,(0: N/2 - 1)/N$ and *fs* is the signal sampling frequency.

### 4.3 Fault Classification

In this work, one-vs-one MCSVM is used to diagnose the gearbox faults. The classifier is trained and tested with 5-fold cross validation by taking gaussian kernel function with optimized kernel scale. Four cases are studied by taking different combination of features set for raw signal and CEEO signal. Table 3 shows the confusion matrix obtained from one-vs-one MCSVM classifier for raw signal and CEEO signal with different types of features set. The diagonal elements show the correctly classified instances and the rest show the misclassified instances. From Table 3, it is clearly observed that the CEEO with combined features (time domain + frequency domain) set have minimum (1) misclassified instances as compared to the other cases. Table 4 shows the overall classification accuracy obtained by the classifier with different features set. For raw signal, the overall accuracy is 85.19 % and 88.89 % with time domain features and combined features set. While for CEEO signal, the overall accuracy is 88.89 % and 96.90 % with time domain features and combined features set, respectively. This comparison shows that the inclusion frequency domain features with time domain features helps to improve the classification accuracy of the classifier and it happened because of the properties claim by the CEEO. The comparison results clearly illustrates the effectiveness of the CEEO data pre-processing over raw signal.

Table 3. Confusion matrix comparison

| Response | Types of features | | Healthy | Chipped | Missing |
|---|---|---|---|---|---|
| (a) Raw signal | Time domain | Healthy | 8 | 1 | 0 |
| | | Chipped | 2 | 7 | 0 |
| | | Missing | 1 | 0 | 8 |
| (b) Raw signal | Time domain + Frequency domain | Healthy | 7 | 2 | 0 |
| | | Chipped | 1 | 8 | 0 |
| | | Missing | 0 | 0 | 9 |
| (c) CEEO signal | Time domain | Healthy | 8 | 1 | 0 |
| | | Chipped | 1 | 8 | 0 |
| | | Missing | 0 | 1 | 8 |
| (d) CEEO signal | Time domain + Frequency domain | Healthy | 9 | 0 | 0 |
| | | Chipped | 0 | 9 | 0 |
| | | Missing | 1 | 0 | 8 |

Table 4. Classification accuracy comparison.

| Response | Types of features | Accuracy |
|---|---|---|
| Raw signal | Time domain | 85.19 % |
| Raw signal | Time domain + Frequency domain | 88.89 % |
| CEEO signal | Time domain | 88.89 % |
| CEEO signal | Time domain + Frequency domain | 96.90 % |

## 5      Conclusion and Discussion

This paper proposed a technique for gearbox fault diagnosis in which a nonparametric data pre-processing (CEEO) was integrated with the one-vs-one MCSVM classifier. The proposed technique has capability to work well under non-stationary elements. The following conclusion are drawn:
- CEEO successfully preserve the characteristics frequencies inside the signal by improving the SIR and SNR value of the signal that helps to improve the quality of feature extraction.
- CEEO has cope up the limitations of previously used data pre-processing technique.
- Comparison results show the effectiveness of CEEO.
- The proposed technique has overall classification accuracy of 96.90 %.

Moreover, the proposed technique can be utilised for the online fault diagnosis of rotary machines and speed up the fault diagnosis process because this technique doesn't require any prior information about the system. In future research, this technique will be modified for fault diagnosis of multi sensor datasets that will take from different parts of gearbox.

## References

1. Sharma, V., Parey, A.:  A Review of Gear Fault Diagnosis Using Various Condition Indicators. Procedia Engineering 144, 253-263 (2016).
2. Leaman, F., Vicuña, C., Clausen, E.: A Review of Gear Fault Diagnosis of Planetary Gearboxes Using Acoustic Emissions. Acoustics Australia *49*(2), 265-272 (2021).
3. Liu, W., Gu, H., Gao, Q., Zhang, Y.: A review on wind turbines gearbox fault diagnosis methods. Journal Of Vibroengineering *23*(1), 26-43 (2021).
4. Kumar, V., Kumar, A., Kumar, S., Sarangi, S.: TVMS calculation and dynamic analysis of carburized spur gear pair. Mechanical Systems and Signal Processing, 166, 108436 (2022).
5. Jin, X., Cheng, F., Peng, Y., Qiao, W., Qu, L.: Drivetrain Gearbox Fault Diagnosis: Vibration- and Current-Based Approaches. IEEE Industry Applications Magazine *24*(6), 56-66 (2018).
6. Kumar, V., Rai, A., Mukherjee, S., Sarangi, S.: A Lagrangian approach for the electromechanical model of single-stage spur gear with tooth root cracks. Engineering Failure Analysis 129, 105662 (2021).
7. Li, C., Sanchez, R., Zurita, G., Cerrada, M., Cabrera, D., Vásquez, R.: Gearbox fault diagnosis based on deep random forest fusion of acoustic and vibratory signals. Mechanical Systems and Signal Processing *76-77*, 283-293 (2016).
8. Caesarendra, W., Tjahjowidodo, T.: A Review of Feature Extraction Methods in Vibration-Based Condition Monitoring and Its Application for Degradation Trend Estimation of Low-Speed Slew Bearing. Machines *5*(4), 21 (2017).
9. Antoniadou, I., Manson, G., Staszewski, W., Barszcz, T., Worden, K.: A time–frequency analysis approach for condition monitoring of a wind turbine gearbox under varying load conditions. Mechanical Systems And Signal Processing *64-65*, 188-216 (2015).

10. Mukherjee, S., Kaushal, R., Kumar, V., Sarangi, S.: A Novel Approach of Gearbox Fault Diagnosis by Using Time Synchronous Averaging and J48 Algorithm. In Advances in Electromechanical Technologies, pp. 927-935. Springer, Singapore (2021).
11. Mukherjee, S., Kumar, V., Sarangi, S., Bera. T.K.: Gearbox Fault Diagnosis using Advanced Computational Intelligence. Procedia Computer Science 167, 1594-1603 (2020).
12. Liang, M., Faghidi, H.: Intelligent bearing fault detection by enhanced energy operator. Expert Systems With Applications *41*(16), 7223-7234 (2014).
13. Kumar, V., Verma, A. K., Sarangi, S.: Fault Diagnosis of Single-Stage Bevel Gearbox by Energy Operator and J48 Algorithm. In Advances in Condition Monitoring and Structural Health Monitoring, pp. 231-239. Springer, Singapore (2021).
14. Mayo, E., Liang, M., Baddour, N.: Gear Fault Detection With the Energy Operator and its Variants. Volume 8: 28Th Conference On Mechanical Vibration And Noise (2016).
15. Vapnik, Vladimir. The nature of statistical learning theory. Springer science & business media, (1999).
16. Hsu, Chih-Wei, Chih-Jen Lin.: A comparison of methods for multiclass support vector machines. IEEE transactions on Neural Networks 13(2), 415-425 (2002).
17. Liu, R., Yang, B., Zio, E., Chen, X.: Artificial intelligence for fault diagnosis of rotating machinery: A review. Mechanical Systems And Signal Processing *108*, 33-47 (2018).
18. Asr, M. Y.; Ettefagh, M. M.; Hassannejad, R.; Razavi, S. N.: Diagnosis of combined faults in Rotary Machinery by Non-Naive Bayesian approach. Mechanical Systems and Signal Processing 85, 56–70 (2017)